\documentclass[aps,prb,showpacs,twocolumn,superscriptaddress]{revtex4}
\usepackage{amsmath}
\usepackage{amssymb}
\usepackage{graphicx}
\begin{document}

\title{Ferromagnetic tendency at the surface of CE charge-ordered manganites}
\author{Shuai Dong}
\affiliation{Department of Physics and Astronomy, University of Tennessee, Knoxville, TN 37996, USA}
\affiliation{Materials Science and Technology Division, Oak Ridge National Laboratory, Oak Ridge, TN 32831, USA}
\affiliation{Nanjing National Laboratory of Microstructures, Nanjing University, Nanjing 210093, China}
\author{Rong Yu}
\author{Seiji Yunoki}
\affiliation{Department of Physics and Astronomy, University of Tennessee, Knoxville, TN 37996, USA}
\affiliation{Materials Science and Technology Division, Oak Ridge National Laboratory, Oak Ridge, TN 32831, USA}
\author{J.-M. Liu}
\affiliation{Nanjing National Laboratory of Microstructures, Nanjing University, Nanjing 210093, China}
\author{Elbio Dagotto}
\affiliation{Department of Physics and Astronomy, University of Tennessee, Knoxville, TN 37996, USA}
\affiliation{Materials Science and Technology Division, Oak Ridge National Laboratory, Oak Ridge, TN 32831, USA}
\date{\today}

\begin{abstract}
Most previous investigations have shown that the surface of a ferromagnetic material may have antiferromagnetic tendencies. 
However, experimentally the opposite effect has been recently observed: ferromagnetism appears in some nano-sized manganites with a composition such that the antiferromagnetic charge-ordered CE state is observed in the bulk. A possible origin is the development of ferromagnetic correlations at the surface of these small systems. To clarify these puzzling experimental observations, we have studied the two-orbital double-exchange model near half-doping $n$=0.5, using open boundary conditions to simulate the surface of either bulk or nano-sized manganites. Considering the enhancement of surface charge density due to a possible $A$O termination ($A$ = trivalent/divalent ion composite, O = oxygen), an unexpected surface phase-separated state
emerges when the model is studied using Monte Carlo techniques on small clusters. This tendency suppresses the CE charge ordering and produces a weak ferromagnetic signal that could explain the experimental observations.
\end{abstract}
\maketitle

\section{Introduction}
Perovskite manganites, with a general formula $A$MnO$_{3}$, where $A$ is the composite of trivalent rare-earth elements and divalent alkaline-earth elements, have attracted considerable attention since the discovery of the colossal magnetoresistance (CMR) effect.\cite{Jin:Sci} Both experimental and theoretical studies in the last decade have unveiled a plethora of phases in manganite compounds, with very different macroscopic properties but very similar energies.\cite{Dagotto:Bok,Dagotto:Prp,Dagotto:Njp,Burgy:Prl} The CMR and colossal electroresistance (CER)\cite{Asamitsu:Nat} effects, which correspond to the observation of drastic nonlinear responses of the manganites to external stimulations, can be understood as a result of the intense competition between the ferromagnetic (FM) metallic phase and antiferromagnetic (AFM) charge-ordered (CO) insulating phases.\cite{Sen:Prl,Yu:Prb,Dong:Prb07,Dong:Jpcm07}

This phase competition is not only sensitive to applied external fields but also to the geometric and chemical environments of the surface of the system under study. In nano-sized materials there is a high surface to volume ratio and, as a consequence, the surface effects play a crucial role. This influence of the surface has been observed in a series of recent experiments. On one hand, in materials where a FM state is stabilized in the bulk, an AFM or spin-glass surface state is found.\cite{Zhang:Jap,Zhu:Apl,Dey:Apl,Moreira:Pb,Zhu:Prb} On the other hand, FM tendencies at the surface of nano-sized manganites presenting AFM/CO bulk order have also been observed.\cite{Biswas:Jap,Rao:Apl,Rao:Prb,Lu:Apl,Zhang:Prb,Sarkar:Prb,Zhang:Jmc,Markovich:Prb,Sarkar:Apl,Biswas:Jap08,Lu:Jap,Biswas:Apl} The first tendency toward surface AFM ordering usually appears in nano-sized FM or ferrimagnetics materials,\cite{Nogues:Prp} and can be understood within the following naive picture. Generally, in strongly correlated electron materials, the charge conducting properties are determined by the ratio $U/W$, where $U$ is the Hubbard Coulomb repulsion and $W$ is the electron hopping bandwidth. At the surface, the bandwidth $W$ will be suppressed due to the reduction in dimensionality, while the on-site $U$ will not change. Therefore, the enhancement of $U/W$ at the surface could prefer an insulating surface over a metallic one. The insulating phase in manganites is usually AFM. However, finding a model and rational for the other tendency found experimentally, namely a FM tendency at the surfaces of CE manganites, is not straightforward. A theoretical model that can explain the FM tendency at the surface should be able to consider the following four experimental signatures. First, the CO phase is significantly weakened. The CO transition peak in the magnetization ($M$) \textit{vs.} temperature ($T$) curve is suppressed until it completely disappears with decreasing the size of the manganite systems.\cite{Biswas:Jap,Rao:Apl,Rao:Prb,Lu:Apl,Zhang:Prb,Sarkar:Prb} Second, $M$ is enhanced at low $T$ and a FM-type hysteresis loop is observed.\cite{Rao:Prb,Lu:Apl,Zhang:Prb,Sarkar:Prb,Zhang:Jmc,Markovich:Prb,Sarkar:Apl} Third, the exchange bias effect emerges, indicating a coupling between phases with different spin orders.\cite{Rao:Prb,Markovich:Prb} And finally, the measurement of magnetocaloric properties and low-$T$ specific heat also suggest the existence of a FM contribution.\cite{Biswas:Jap08,Lu:Jap,Biswas:Apl}

Theoretically, previous investigations have mainly paid attention to the surface effects of FM manganites \cite{Calderon:Prb,Zenia:Prb,Zenia:Njp} whereas the study of the surface of AFM manganites was rare. To understand these phenomena, it is reasonable to partition the nano-sized (with typical scales $10\sim10^{2}$ nm) system into two regions: an inner core and a surface shell. The physical properties of the inner core should be comparable to those of the bulk material. In contrast, the properties of the surface shell can be different from those of the bulk. The lower coordination number at the surface, which effectively reduces the superexchange coupling, may give rise to a FM tendency at that surface. Following this idea, in previous investigations a core-shell model was proposed assuming an AFM core wrapped by a fully FM surface shell.\cite{Dong:Apl07} This model fits some experimental results well, but in this scenario it is already established from the start the nature of the phases at both the core and surface. As a consequence, it is too phenomenological to better understand the true physical origin of the AFM-FM transition at the surface. In particular, the assumption of a fully FM shell conflicts with some experimental results: due to the weak value of the magnetization $M$, the calculated thickness of the FM shell can be even thinner than one ``molecular'' layer, which indicates that assuming a fully developed FM spin order at the surface is not correct. Thus, theoretical studies using realistic microscopic Hamiltonians and unbiased assumptions about the surface are necessary to clarify the experimental observations found at the surface of nano-sized AFM/CO manganites.

In this paper, the core-shell model is incorporated into a two-orbital Hamiltonian for manganites. We show that the unscreened Coulomb interactions lead to an increase of electron density on the surface. Monte Carlo simulations reveal that this increase in the density drives the surface layer from an AFM/CO state to an unexpected phase-separated state, as opposed to a fully developed FM state. This surface phase-separated state exhibits clear FM signatures, but they are weak, compatible with the experimental observations.

\section{Model}
To better understand the physics at the surface, here we consider a two-orbital model Hamiltonian for manganites that includes both finite superexchange coupling and the effect of Jahn-Teller phonons. As a well-accepted approximation for manganite models, we consider the limit of infinite Hund coupling. The Hamiltonian reads:
\begin{eqnarray}
\nonumber H&=&-\sum_{<ij>}^{\alpha\beta}t_{\textbf{r}}^{\alpha\beta}\Omega_{ij}c_{i\alpha}^{\dagger}c_{j\beta}+J_{AF}\sum_{<ij>}\textbf{S}_{i}\cdot \textbf{S}_{j}\\
&&\nonumber+\sum_{i}(\epsilon_{i}-\mu)n_{i}+\lambda\sum_{i}(Q_{1i}n_{i}+Q_{2i}\tau_{xi}+Q_{3i}\tau_{zi})\\
&&+\frac{1}{2}\sum_{i}(2Q_{1i}^2+Q_{2i}^2+Q_{3i}^2).
\end{eqnarray}
Here, the first term is the two-orbital double exchange interaction. $\alpha$ and $\beta$ denote the two Mn $e_{g}$-orbitals $a$ ($d_{x^2-y^2}$) and $b$ ($d_{3z^2-r^2}$). $c_{ia}$ ($c_{ia}^{\dagger}$) annihilates (creates) an $e_{g}$ electron in orbital $a$ of site $i$ with its spin parallel to the localized $t_{2g}$ spin $\textbf{S}_{i}$. The hopping direction is
denoted by \textbf{r}. As discussed in previous literature,\cite{Dagotto:Bok,Dagotto:Prp} the hopping amplitudes are $t_{x}^{aa}=t_{y}^{aa}=3t_{x}^{bb}=3t_{y}^{bb}=3/4$, $t_{y}^{ab}=t_{y}^{ba}=-t_{x}^{ab}=-t_{x}^{ba}=\sqrt{3}/4$, $t_{z}^{aa}=t_{z}^{ab}=t_{z}^{ba}=0$ and $t_{z}^{bb}=1$ (energy unit). The infinite Hund coupling generates the factor
$\Omega_{ij}=\cos(\theta_{i}/2)\cos(\theta_{j}/2)+\sin(\theta_{i}/2)\sin(\theta_{j}/2)\exp[-i(\phi_{i}-\phi_{j})]$, where $\theta$ and $\phi$ are the angles of the $t_{2g}$ spins in spherical coordinates. The second term is the superexchange interaction between nearest-neighbor (NN) $t_{2g}$ spins. In the third term, $\mu$ is the chemical potential. $\epsilon_{i}$ corresponds to a site dependent Coulomb potential. The origin and relevance of this term will be discussed in detail later in this section. $n_i$ is the $e_{g}$ charge-density at site $i$. The fourth term stands for the electron-phonon coupling. The $Q$s are phonons corresponding to Jahn-Teller modes ($Q_{2}$ and $Q_{3}$) and the breathing mode ($Q_{1}$). $\tau$ is the orbital pseudospin operator,  giving $\tau_{x}=c_{a}^{\dagger}c_{b}+c_{b}^{\dagger}c_{a}$ and $\tau_{z}=c_{a}^{\dagger}c_{a}-c_{b}^{\dagger}c_{b}$. The last term is the elastic energy of the phonons. For simplicity, we have already assumed in this model that both the $t_{2g}$ spins and the phononic degrees of freedom are classical variables. The above described Hamiltonian is solved via a combination of exact diagonalization and Monte Carlo (MC) techniques: classical $t_{2g}$ spins and phonons evolve following the MC procedure; and at each MC step, the fermionic sector of the Hamiltonian is numerically exactly diagonalized. The first $10^{4}$ MC steps are used for thermal equilibrium and another $6\times10^3$ MC steps are used for measurement. More details of the Hamiltonian and MC technique can be found in Refs.~2,3.

\begin{figure}[t]
\includegraphics[width=220pt]{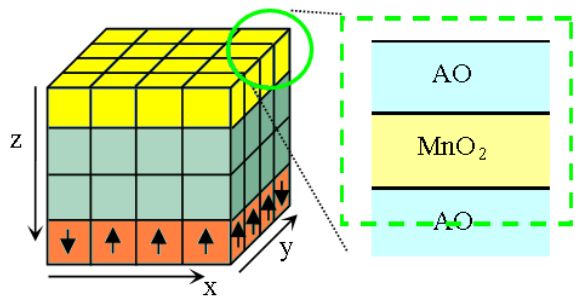}
\caption{(Color online) Sketch of the model and geometry used in this investigation. \textit{Left}: a cubic lattice with one open surface (yellow layer). All spins in the bottom layer (orange) are frozen to the CE-type AFM pattern. \textit{Right}: the chemical unit taken from the surface layer of the cube at the left. The $A$O sheet termination is considered here. Thus, the surface formula is $A_{1.5}$MnO$_{3.5}$.}
\end{figure}

Since a FM tendency is often found in half-doped \cite{Rao:Apl,Rao:Prb,Biswas:Jap,Sarkar:Apl} or nearly half-doped\cite{Lu:Apl,Lu:Jap} nano-sized manganites, and since a CE-type AFM/CO phase usually appears in half-doped narrow band manganites,\cite{Wollan:Pr} we solve the two-orbital model using densities corresponding to half-doped systems and for couplings where a bulk CE phase exists. This can be done by tuning the couplings ($J_{AF}$, $\lambda$) to be within the CE regime of the phase diagram obtained in previous studies.\cite{Yunoki:Prl,Pradhan:Prl} Although only one set ($J_{AF}$, $\lambda$) will be investigated in the studies described below, we believe our qualitative conclusions remain valid for other choices of parameter sets, as long as they are within the the CE region of the phase diagram.

Although there are several possible surface directions in real cases, here only the simplest case, i.e., the (001) surface, will be considered, similarly as in most former theoretical
investigations in this context. \cite{Calderon:Prb,Zenia:Prb,Zenia:Njp,Fang:Jpsp} Strictly speaking, the surface problem should be considered on a three-dimensional half-infinite cubic lattice, e.g. infinite in the $x$ and $y$ directions, but semi-infinite in the $z$ direction. In practice, to address this problem numerically the above two-orbital model Hamiltonian is studied using a $L\times L\times L$ cubic lattice, as shown in Fig.~1. Periodic boundary conditions (PBC) are applied in both the $x$ and $y$ directions. Following the core-shell-model idea discussed in the previous section, in the $z$ direction the spins in the bottom layer ($z=L$) are here \textit{fixed} so that they have the same CE-type AFM pattern as in the bulk. By this procedure, the properties of the half-infinite core are encoded in the bottom layer in our model. Hence, the rest $L-1$ layers are to be considered as the outer shell. To better justify the above assumption, $L$ must be large enough that surface effects are limited within the outer $L-1$ shell layers. Our results below suggest that in practice $L=4$ is enough for these purposes. This is fortunate, since the numerical studies described here are rather CPU time consuming.

For the surface layer ($z=1$), we have to take into account the ``termination'' procedure. In this paper, a clean-limit $A$O sheet is considered as the termination, which makes the ``molecular composition'' of the surface layer to be $A_{1.5}$MnO$_{3.5}$. Therefore, the outmost $A$O sheet transfers an extra $0.25$ electron per site to the nearby Mn cations, and it is positively polarized. The polarized $A$O sheet introduces an unscreened Coulomb attraction on the surface. As a result, those extra electrons must be localized near the surface. The effect of this surface Coulomb attraction has been taken into account here via an effective negative potential near the surface: $\epsilon_{i}=V<0$ when $i$ belongs to the surface layer and $\epsilon_{i}=0$ otherwise. This may be justified because of the typical short Thomas-Fermi-like screening length found in many of these materials. The boundary condition on the surface layer is then set to be open. The consideration of the $A$O sheet has another important effect: it keeps the oxygen octahedrons complete for the outmost Mn cations. Therefore, the phonon modes do not need to be changed, even for the surface layer. For the phonons, PBC are used in all directions for simplicity. This will not affect much the central physics discussed in this manuscript since the oxygen displacements along the $z$ axis are negligible in both the FM and CE type AFM phases.\cite{note1} In addition to the $A$O termination, the other possible choice for the surface is a clean-limit MnO$_{2}$ sheet as the termination, but this will not be considered in our current model. However, the possible effects of this alternative termination will also be discussed in the next section.

To contrast results with those of the open surface case, the model will also be studied following assumptions that address the bulk material: this corresponds to using PBC in all directions, for both spins and phonons, and without freezing spins anywhere. Both simulations are performed at $T=0.02$ (experimentally, it corresponds to $50\sim100$ K). To simulate the bulk material, we set $\epsilon_{i}=0$ for all sites. The parameters $J_{AF}=0.1$ and $\lambda=1.2$ are used to obtain a stable CE phase with average density $<n_i>=0.5$. Note that it is well known that the CE phase is stable over a broad range of couplings of the half-doped manganites, thus this selection of parameters should not be considered arbitrary or fine tuned. The stability of the CE phase is also confirmed in the present simulations by analyzing the spin and charge structure factors, which will be discussed in detail in the next section. As observed in Fig.~2(a), this corresponds to choosing the chemical potential in the window $-1.1<\mu<-0.95$. These parameters will also be adopted in our subsequent simulation of the open surface model.

To obtain reasonable results from the simulation of the open surface model, we have to set the Coulomb potential to an appropriate value. Actually, an optimal value of $V$ exists in order to fulfill the following three criteria: first, the average density per site of the entire $L\times L\times L$ system $<n_i>$ should be equal to $(0.5+0.25/L)=0.5625$ to keep the charge neutral; second, the chemical potential should remain within ($-1.1$, $-0.95$), which is required to have the CE phase stable far from the surface; third, the charge density in the bottom layer should be very close to $0.50$ to match the frozen CE type. In order to find the optimal $V$, we tested several values from $0$ to $-0.6$, stepped by $-0.1$. In this range, $V=-0.4$ was found to be a proper parameter at $\mu=-1$.  Thus, $V=-0.4$ was the value adopted in the 
following simulations.

\begin{figure}[b]
\includegraphics[width=220pt]{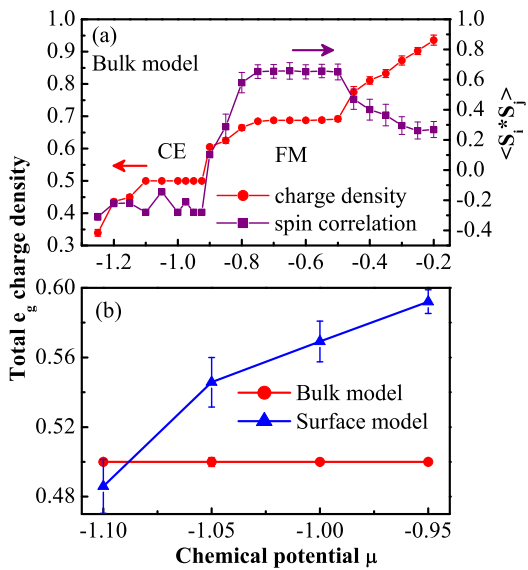}
\caption{(Color online) The total 
$e_{g}$ charge-density as a function of the chemical potential using $L=4$. (a) is for the case of the bulk material model. The corresponding NN spin-correlation is also shown. There is a $<n_i>=0.50$ plateau with the NN spin-correlation $\sim-0.3$ in
the range $-1.1<\mu<-0.95$, indicating a stable CE phase. (b) contrasts the results for the open-surface and the bulk material models near half-doping.}
\end{figure}

\section{Results and discussion}
Using as initial configuration a perfect CE spin pattern \cite{note2} and random phonons, the MC simulation was carried out  under the assumption $V=-0.4$ for the open-surface model. The averaged $e_{g}$ charge-density as a function of $\mu$ is shown in Fig.~2(b), allowing us to compare the results for the bulk material model with those of the open surface. An interesting result is that the $n$=$0.5$ plateau in the bulk model, which corresponds to the stable CE phase, disappears in the open-surface model. The charge density increases with increasing chemical potential but it has large fluctuations (see error bars), suggesting that the system presents a  strong competition between phases with different densities. In the following, we will focus on the properties at $\mu=-1$, with the average total charge density close to $0.5625$.\cite{note3}

To understand the origin of the different average charge density between bulk and open-surface models shown in Fig.~2(b), it is important to analyze the $e_{g}$ charge-density at each layer. The results are shown in Fig.~3(a). The charge density is almost exactly $0.50$ at the bottom layer and fluctuates around $0.50$ in the two middle layers. The most prominent change occurs at the surface layer, where the charge density increases to $0.75$, due to the presence of the Coulombic term. As expected, the extra $0.25$ electron from the outmost $A$O sheet is mainly located in the first Mn sheet, which offsets the Coulomb interaction arising from the outmost $A$O sheet for the Mn cations of the second layer. This is consistent with $ab$-$initio$ calculations that show that for the $A$O termination the uncompensated electrons are accumulated mainly at the surface.\cite{Fang:Jpsp} This result is also consistent with the assumption that $\epsilon_{i}$ is nonzero only for the first layer.

\begin{figure}[t]
\includegraphics[width=220pt]{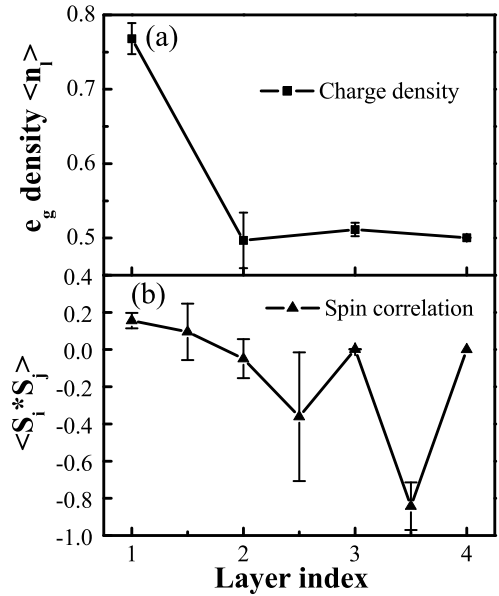}
\caption{(a) The $e_{g}$ charge-density for each layer (counted from the surface). (b) The corresponding averaged NN spin-correlation,  within (integer index) and between (half-integer index) layers.}
\end{figure}

It should be noted that the density $0.75$ usually corresponds to a FM phase in bulk manganites, as indicated in Fig.~2(a). If the bulk phase diagram remains valid at the surface, we would expect a FM state there. Then, the FM tendency could be naturally explained as a result of a density-driven transition. To verify this possible FM tendency, the average NN spin-correlation as a function of layer index is shown in Fig.~3(b). Starting from the bottom layer ($L=4$), we find that the in-plane NN spin correlations are almost zero for layer $L=4$ and $L=3$, implying that the FM and AFM links have almost the same population. The inter-layer correlation between these two layers is close to $-1$, indicating a fully AFM connection between the two layers. These are consistent with the picture that a CE AFM state is stabilized in these two layers, given that the charge density is about $0.50$. Interestingly, we do see an increase of both the in-plane and inter-layer NN spin-correlations as we approach the surface. Both the correlations in the first layer and between the first and second layers take positive values, and they display a clear FM tendency at the surface layer. But the positive value for the in-plane correlation is rather small ($\sim0.15$) at the surface layer, which suggests that the state is only \textit{partially} FM. Therefore, the idea of a density-driven transition is too simplistic and not quite correct. This already shows an interesting conclusion of our research: the phase diagram at the surface cannot be obtained by merely analyzing the bulk phase diagram at the appropriate charge density, but a special investigation is needed to clarify the surface's properties.

\begin{figure}[b]
\includegraphics[width=240pt]{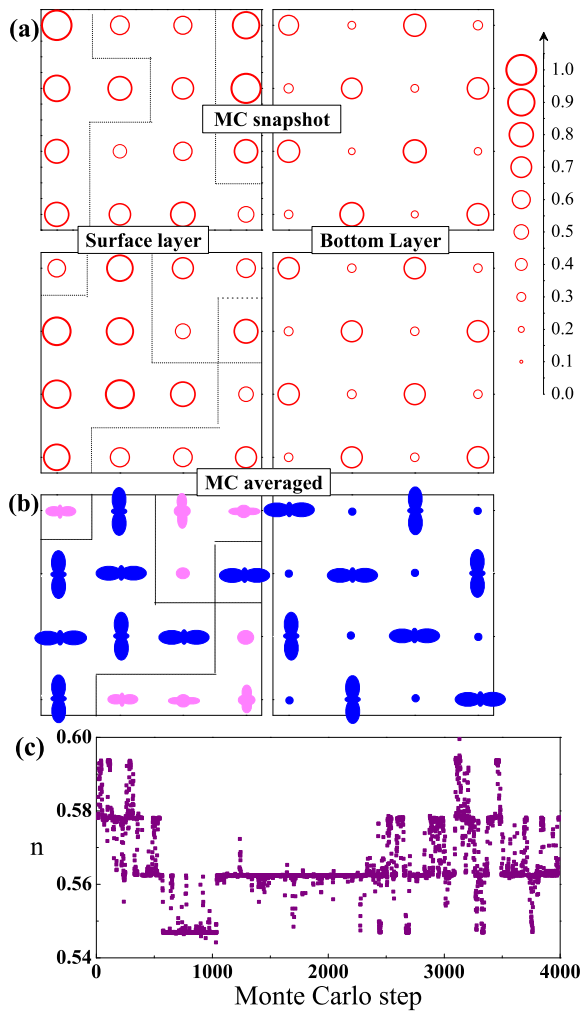}
\caption{(Color online) (a) The $e_{g}$ charge-distribution corresponding to the surface layer (left) and bottom layer (right). Here are shown both a typical MC ``snapshot'' (upper panels) and MC averaged results (lower panel, averaged over $6,000$ MC steps). The size of the circles is in proportion to the local charge density $n_{i}$. For the surface layer, the dotted lines separate the high density ($n_{i}>0.75$) and low density ($n_{i}<0.75$) regions. (b) Sketch of the orbital occupation  (based on the MC-averaged $\tau_{xi}$ and $\tau_{zi}$) corresponding to the above-described MC-averaged charge distribution. (c) MC-time evolution after thermal equilibrium of the $e_{g}$ charge-density (average value for all the sites). The frequent tunneling events are prominent among the possible densities, indicating tendencies toward electronic phase separation, which is a first-order transition when $\mu$ is varied.}
\end{figure}

To visualize the nature of the surface state more explicitly, we study the distribution of local charge density on the surface layer. In a FM phase, the distribution of local charge density $n_{i}$ is approximately uniform, in contrast to the charge disproportion typical of an AFM/CO phase.\cite{Dong:Prb} The distribution of local charge density on the surface layer of the model studied here is presented in Fig.~4(a), and the result at the bottom layer is also shown in the same figure for contrast. From the regular charge pattern, it is clear that the CE charge ordering in the bottom layer is very stable. Since the CE spin pattern is fixed at the bottom layer, the charge disproportion between large-density sites (bridge-sites of the zig-zag chains) and smalle-density sites (corner-sites of the zig-zag chains) will not be smeared by MC average. However, for the surface layer the charge distribution is not uniform, and it is not regularly distributed. In some sites, the densities are large and close to $1$, but in other sites the densities can be as low as approximately $0.5$. This inhomogeneous distribution persists prominently even in the MC-averaged result, which rules out the possibility of observing a non-uniform charge state due to thermal fluctuations. This inhomogeneity can also be confirmed from the orbital occupation, as shown in Fig. 4(b).  In contrast to the distinct CE type orbital-ordering in the bottom layer, the orbital distribution in the surface layer shows two regions corresponding to charge inhomogeneity. The high-density region shows an orbital ordering similar to the case in undoped manganites, in contrast to the low-density region which shows the orbital disorder similar to the case in FM manganites. In addition, the MC-time evolution of the $e_{g}$ charge-density is presented in Fig.~4(c), showing that the tunneling events are prominent among several possible densities. These tunneling events are characteristic of a first-order phase transition varying $\mu$, rather than standard thermal fluctuations.\cite{Dagotto:Prb} Therefore, the inhomogeneous charge distribution at the surface should be attributed to tendencies in the model toward nanoscale electronic phase separation,\cite{Dagotto:Bok,Dagotto:Prp} similarly as those observed in bulk simulations in other regions of parameter space.

\begin{figure}[b]
\includegraphics[width=220pt]{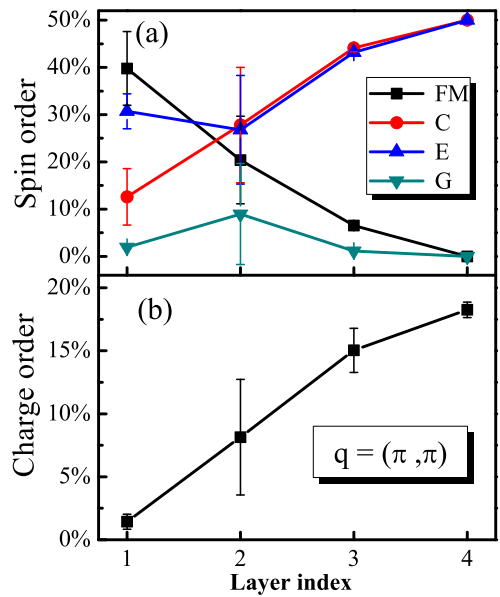}
\caption{(Color online) (a) The spin structure factor values for several spin orders, showing the reduction in the tendency toward a CE state as the surface is reached. At this surface, the FM and E phase tendencies are dominant. (b) The charge structure factor at ($\pi$,$\pi$) as a function of the layer index, showing the reduction of the CE-phase staggered charge-order tendencies as the surface is reached.}
\end{figure}

To quantitatively reveal the competing phases that appear at the surface, the spin structure factor $S(\textbf{q})$ for each layer is calculated via a Fourier transformation of the spin correlation function:\cite{Sen:Prb}
\begin{equation}
S(\textbf{q})=\frac{1}{L^{4}}\sum_{ij}\textbf{S}_{i}\cdot\textbf{S}_{j}e^{i\textbf{q}\cdot(\textbf{r}_{i}-\textbf{r}_{j})}.
\end{equation}
By monitoring the $\textbf{q}$ dependent spin structure factor, we may detect possible weak spin-order signals in the complicated real-space spin pattern\cite{Pradhan:Prl,Kumar:Prl08} because each spin order corresponds to a unique set of characteristic $\textbf{q}$ vectors.\cite{note4} In Fig.~5(a), the $S(\textbf{q})$s for some possible components (summed over all characteristic $\textbf{q}$ vectors for each spin order) are shown layer by layer. For the bottom layer, which is frozen into a CE spin pattern, both the C and E components contribute $50\%$, respectively, as expected. The CE phase tendencies become gradually weaker with decreasing layer index, indicated by the decreasing values and large fluctuations of the corresponding $S(\textbf{q})$s. At the surface, the C component is very weak ($\sim10\%$) and the dominating components here are FM ($\sim40\%$) and E ($\sim30\%$). In this case, the E value does not match the C value anymore because the site regions where $n_{i}\approx1$ can also contribute to the E type order.\cite{Hotta:Prl}

In addition to the $S(q)$ spin structure factor, the charge structure factor $C(\textbf{q})$ in each layer is also calculated to characterize the charge ordering:
\begin{equation}
C(\textbf{q})=\frac{4}{L^{4}}\sum_{ij}(n_{i}-n_{l})(n_{j}-n_{l})e^{i\textbf{q}\cdot(\textbf{r}_{i}-\textbf{r}_{j})},
\end{equation}
where $n_{l}$ is the average density of each layer. Here, only $C$($\pi$,$\pi$) is shown in Fig.~5(b) since all other components are very close to zero. For the CE phase at the bottom layer, $C$($\pi$,$\pi$) is about $18\%$, a result consistent with the expected charge-ordering with charge disproportion $\sim0.4$ (the charge difference between high- and low-density sites). Decreasing the layer index, $C$($\pi$,$\pi$) decreases monotonically as we move toward the surface, until it completely disappears at the surface. This is also straightforward to understand since charge-ordering usually accompanies the AFM phases instead of the FM ones, even in bulk manganites.

In the above simulation, the ground state is a robust CO CE phase by fixing the proper $J_{AF}$ and $\lambda$. Therefore, it is worth to address the other possible cases in real half-doped manganites. Here we will give a brief analysis based on the half-doped phase diagram obtained in previous works.\cite{Yunoki:Prl,Pradhan:Prl}  On one hand, by decreasing $J_{AF}$ only, the ground phase can change from the CE to A-type AFM, then finally to FM phase. This process corresponds to the experimental observed transition from the narrow band manganites to middle band one, then finally to wide band manganties.\cite{Kajimoto:Prb} On the other hand, by decreasing $\lambda$ only, the charge-ordering (or the degree of charge disproportionation) will become weaker and weaker, till completely turn to the FM phase when ($J_{AF}$, $\lambda$) crosses the phase boundary between CE and FM phases. In both cases, FM tendency will be enhanced. In short, our above simulation mainly aims at the family of narrow band manganties whose CO CE phase is stable, e.g. Nd$_{0.5}$Ca$_{0.5}$MnO$_{3}$ and Pr$_{0.5}$Ca$_{0.5}$MnO$_{3}$.

In the simulations described above, only the $A$O sheet termination was considered. But it is necessary to discuss the other choice already mentioned, namely a MnO$_{2}$-sheet termination. In the case of this MnO$_{2}$ termination, the cubic symmetry is lost at the surface due to the breakdown of the oxygen octahedrons. Therefore, the $3d$ energy levels of the Mn cations at the surface are different from those with full oxygen octahedrons. In particular, the energy of the $d_{3z^2-r^2}$ orbital will be lowered substantially. An early model study by Calder{\'o}n \textit{et al} showed that the MnO$_{2}$ termination would generate an AFM surface for FM manganites because the $e_{g}$ density at the surface was enhanced from $\sim0.7$ to $1$.\cite{Calderon:Prb} However, an $ab$-$initio$ calculation by Fang \textit{et al.} showed the reverse result: decreased $e_{g}$ density at the surface by MnO$_{2}$ termination.\cite{Fang:Jpsp} Therefore, whether the MnO$_{2}$ termination can generate the FM tendency, e.g. by enhancing the surface charge density from $0.5$ to $0.75$ as it occurs in the case of the $A$O termination, remains unclear and is an interesting subject of investigations. However, this MnO$_{2}$ termination is far more complex than the case studied here, and beyond the scope of the present work.

It is important to remark that some other extrinsic factors, such as defects of cations/oxygen and recomposition of surface structures, may also affect the physical properties of real manganites. Even qualitatively considering these effects, our model still gives reasonable results. For a crude comparison with experiments, the FM component fraction can be estimated as $40\%$ of the surface layer (Fig.~5(a)). Therefore, to compare with experiments the FM fraction predicted by our study we should use the number $40\%\times$surface/volume (at zero magnetic field and low $T$), where surface/volume should be calculated based on the actual shape and size of the nano-sized clusters used experimentally. By this procedure, our theory agrees with the weak magnetization found in Nd$_{0.5}$Ca$_{0.5}$MnO$_{3}$ nanoparticles (diameter $\sim20$ nm, experimental magnetization $M\sim3$ emu/g $\sim3\%$ of the saturation magnetization $M_{s}$, while in our model our estimation gives $\sim5\%M_{s}$)\cite{Rao:Prb} and Pr$_{0.5}$Ca$_{0.5}$MnO$_{3}$ nanowires (diameter $\sim50$ nm, experimental magnetization $M\sim1$ emu/g $\sim1\%M_{s}$, while our model estimation is $\sim1\%M_{s}$).\cite{Rao:Apl} It should be noted that the phenomenological core-shell model can not explain these very weak magnetizations.\cite{Dong:Apl07} The agreement between our estimates and experimental data suggests that our model at least grasps the main physics of the surface effects in the nano-sized CE-phase manganites.

\section{Conclusion}
In conclusion, we have performed a Monte Carlo study of the CE-type AFM/CO phase in a 3D lattice with an open surface. The $A$O sheet termination leads to the generation of an extra $0.25$ electron-per-site at the surface, here simulated by the introduction of an unscreened Coulomb attraction at that surface. As a result, the charge density on the surface Mn-layer was enhanced from $0.50$ to $\sim0.75$. The charge density $\sim0.75$ usually corresponds to a fully FM phase in bulk manganites. However, within the Monte Carlo simulations for small clusters discussed in this manuscript, the surface was found to have a nontrivial nanoscale electronic phase separated state. At the surface, the charge distribution was found to be inhomogeneous and coexisting with a weak FM spin correlation. The studies of both the charge structure factor and spin structure factor confirmed the suppression of AFM/CO order and  the enhancement of FM order near the surface. Our result is helpful to understand the weak FM tendencies observed in nano-sized AFM/CO manganites. However, clearly these results have to be considered as just a first step toward the understanding of the phase diagram at the surfaces of manganites. Larger clusters and other numerical/analytical techniques should be used to confirm our results and further explore the physics unveiled here. Also, a systematic study varying parameters of the many tendencies expected in the anticipated rich phase diagram of these compounds at the surface should be carried out in future investigations.

\section{Acknowledgment}
We thank W. Plummer, M. J. Calder\'on, S. V. Bhat, and A. Biswas for useful comments. This work was supported by the NSF grant DMR-0706020 and the Division of Materials Science and Engineering, U.S. DOE, under contract with UT-Battelle, LLC. S.D. and J.M.L were supported by the National Key Projects for Basic Research of China (2006CB921802). S.D. was also supported by the China Scholarship Council and the Scientific Research Foundation of Graduate School of Nanjing University.


\begin{thebibliography}{bib}
\bibitem{note1} Usually, the phononic displacements are prominent in CO insulating phases but weak in FM metallic phases. For the CE phase, the $2$-dimensional ($x$-$y$ plane) nature of the spin pattern causes the oxygen displacements along the $z$ direction to be zero. For more details, see Refs.~2,3.
\bibitem{note2} The CE zig-zag chains are arranged in the (001) plane. Therefore, the (001) surface will not break the in-plane zig-zag chains topologically. This choice can avoid the finite size effects along the [100] and [010] directions.
\bibitem{note3} It is very difficult to obtain the exact $n$=$0.5625$ density due to the phase separation tendencies in the grand canonical ensemble. The MC-averaged density at $\mu=-1$ with $V=-0.4$ is $0.569\pm0.012$. This small discrepancy of charge density does not affect the conclusions of our manuscript.
\end{thebibliography}
\end{document}